\documentclass[twocolumn,aps,prb,floatfix,showpacs]{revtex4}

\usepackage{amsmath}
\usepackage[mathscr]{eucal}
\usepackage{graphics}
\usepackage{subfigure}

\begin{document}

\widetext

\title{Effect of Symmetry Distortions on Photoelectron Selection
  Rules and Spectra on
  Bi$_2$Sr$_2$CaCu$_2$O$_{8+\delta}$}

\author{V. Arpiainen, M. Lindroos}

\affiliation{   Institute of Physics, Tampere University of
  Technology, P.O. Box 692, 33101 Tampere, Finland}

\date{\today}

\begin{abstract}
   We derive photoelectron selection rules along the glide plane in
    orthorhombic Bi$_2$Sr$_2$CaCu$_2$O$_{8+\delta}$ (Bi2212). These
    selection rules explain the reversed intensity behavior of the
    shadow and the main band of the material as a natural consequence of the
    variating representation of the final state as a function of
    $\boldsymbol k_\parallel$. Our one-step simulations strongly
    support the structural origin of the shadow band but we also
    introduce a scenario for detecting antiferromagnetic signatures in
    low doping.
\end{abstract}

\maketitle
 Angle Resolved Ultraviolet Photoelectron Spectroscopy (ARUPS) is
 probably the most important tool to study the
electronic structure of complex materials like the high T$_c$ superconductors. 
It has been shown 
that matrix element effects
must be taken into account in interpreting experimental
data\cite{Ban99,Ase03}. Using the one-step model of ARUPS and optical matrix
elements, light has been shed on the strong variation of intensities due
to the matrix element effects.\cite{Lin02,Sah03,Ban05}
However, there are more strict rules dictating intensites in ARUPS, selection rules due to crystal
symmetry.
High resolution angular resolved photoemission  measurements along
high symmetry lines
have been carried out recently to investigate the origin of the shadow bands
in Bi2212 \cite{Man06,Izq05}. 
Based on one-step computations,
it was proposed that the shadow Fermi
surface(FS) in Bi2212 has its origin in the orthorhombic (orth.) structure
of the material.\cite{Man06}  In this study a need for selection
rules emerged.  In the orth. structure a mirror plane is broken into a
glide plane. In this letter we give a
group theoretical analysis for the selection rules along the glide
plane and demonstrate the derived rules with one-step ARUPS
simulations. 
We focus on
Bi2212, but the derived selection rules
may be applied to any cuprate superconductor with a
similar structure.

The space group of the commonly used body-centered tetragonal(tetr.) structure for Bi2212
is number 139 in the International Tables for
Chrystallography.
In x-ray
diffraction experiments
an orth.
structure has been found. \cite{Mil98} In this
structure atoms are moved from their tetr.
positions and the symmetry group is changed.
The non-symmorphic space group of the
orth. structure is the group number 66. 
\begin{figure}[]
  \subfigure[]{
    \resizebox{2.35cm}{!}{\includegraphics{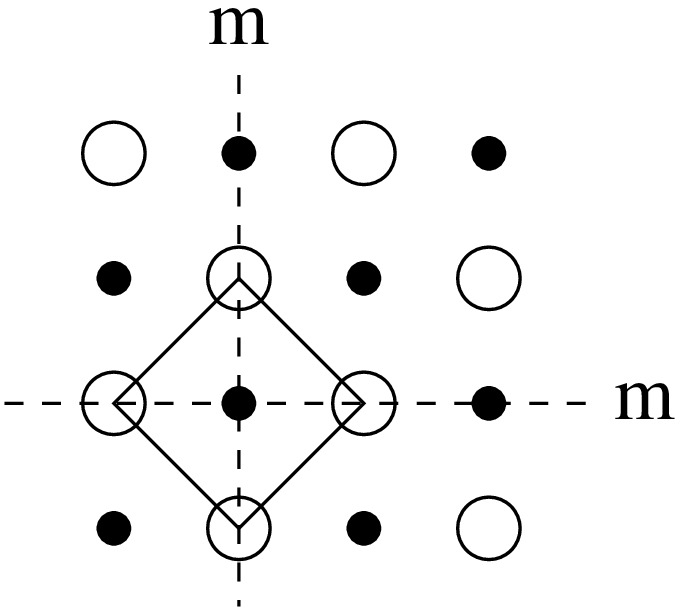}}
    \label{bitetra}
  }
  \subfigure[]{
    \resizebox{2.961cm}{!}{\includegraphics{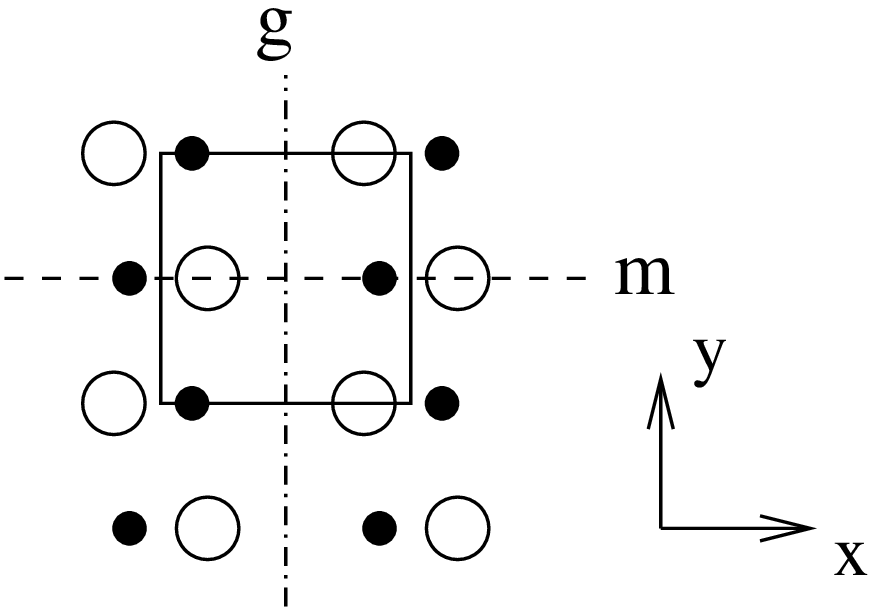}}
    \label{biortho}
}
 \subfigure[]{
    \resizebox{2.4cm}{!}{\includegraphics{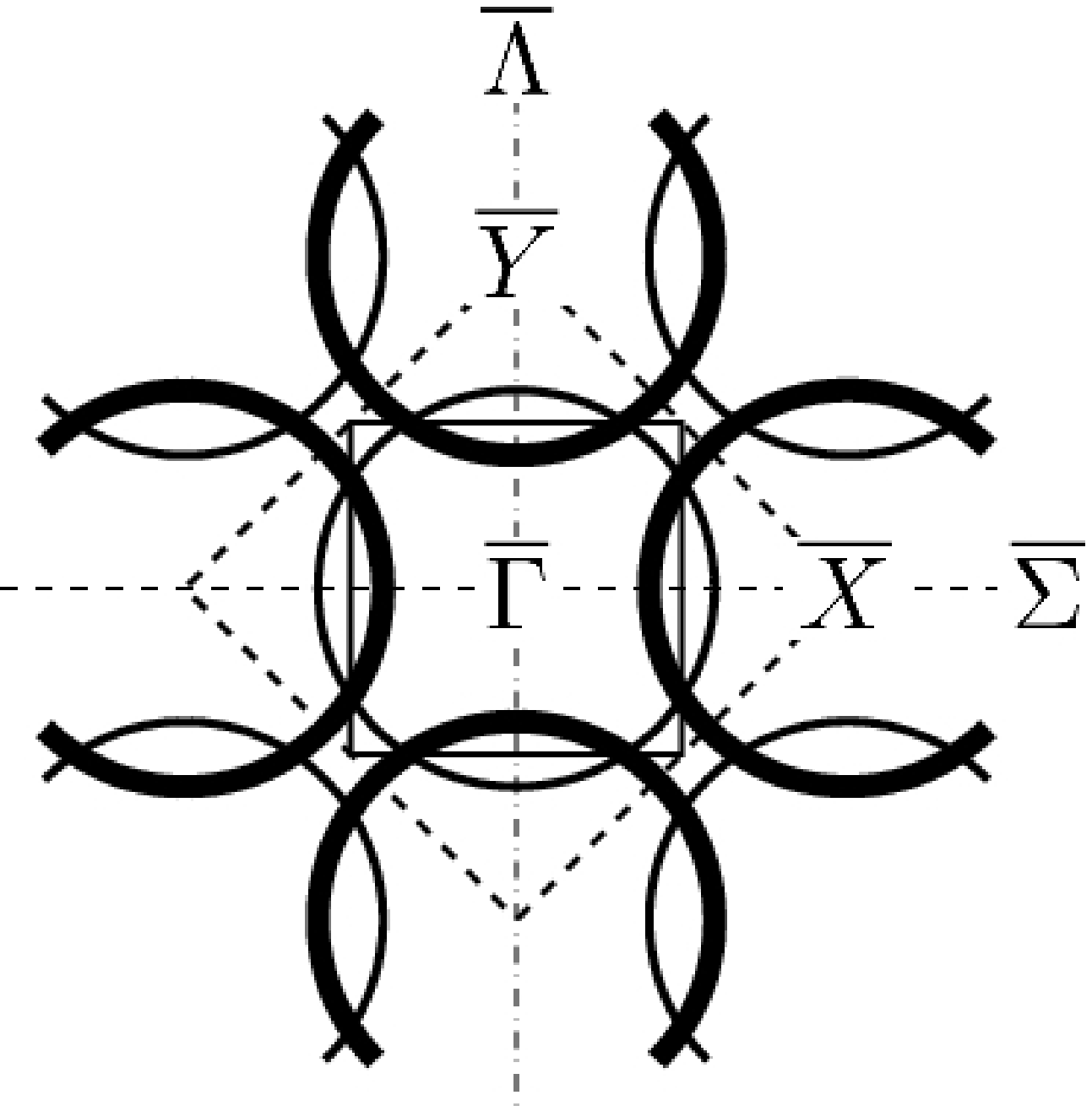}}
    \label{k_kaavio}
}
\caption{Sketch of the positions of atoms in Bi2212. (a) Tetr. Bi-O layer, (b) orth. Bi-O layer, (c) Sketch of the Fermi surface.}
\label{Bilayer}
\end{figure}
The geometric structure of the superconducting cuprates may be described
by a pile of planes of atoms. In the orth. structure the movements of
the atoms from
tetr. positions in the  Cu-O-planes are small. Displacements are
more clear in the Bi-O-plane, which is scetched in
Fig. \ref{Bilayer}.
Fig. \ref{bitetra} represents the tetr.
Bi-O-plane and Fig. \ref{biortho} the orth. Bi-O plane, where
empty circles representing Bi-atoms and filled circles
representing O-atoms are displaced. 
In the tetr. structure there are mirror
planes corresponding to dashed lines in Fig. \ref{bitetra}, but in the orth. structure only the mirror
plane parallel to $x$ axis remains. The mirror plane
parallel to $y$ axis is changed into a
glide plane which is plotted with dash-dotted line in Fig.
\ref{biortho}.  This glide plane corresponds the glide-operation $\{\sigma_{x}|{\bf
  b}/2\}$, a reflection ($\sigma_{x}$ in the $x = 0$ plane) followed by a
translation by half the primitive lattice vector in the
$y$ axis. 

The model FS of Bi2212 (only one sheet is shown for clarity)
is sketched in Fig.
 \ref{k_kaavio}. The figure shows the main band (FS of the
 tetr. structure) as thick circles and
 the shadow band as thin circles. In the figure the first tetr. Brillouin zone is
 approximately plotted with dashed line and the first orth. Brillouin zone
 with solid
line. 2-dimensional $k_{\parallel}$-points of high symmetry $\bar\Gamma$,$\bar X$ and
 $\bar Y$ are also marked.
 The direction from $\bar\Gamma$ to $\bar X$ that is parallel to the mirror plane in
 real space is denoted by $\bar \Sigma$, and the direction from $\bar\Gamma$
 to $\bar Y$
 that is parallel to the glide plane by $\bar \Lambda$.\cite{Fle71}
 These directions are commonly known as the nodal directions.

The formula, based on the Fermi's golden rule, for the photocurrent by
Feibelman and Eastman\cite{Fei74} can be manipulated into the form
\begin{align}
I(E,\boldsymbol k_\parallel) \propto Im( \sum_{ij}
\langle\psi_{f,\boldsymbol k_\parallel}^{(final)}|{\bf A} \cdot {\bf
  p}|\psi_{i,\boldsymbol k_\parallel}^{(initial)}\rangle \nonumber \\
 G_{ij} 
\langle\psi_{j,\boldsymbol k_\parallel}^{(initial)}|{\bf A} \cdot {\bf
  p}|\psi_{f,\boldsymbol k_\parallel}^{(final)}\rangle ),
\label{goldenrule}
\end{align}
where ${\bf A}$ is the vector potential of the
incident photon and $G_{ij}$ is a hole Green function.
Photoemission selection rules
arise from the symmetries of the initial ($\psi_i$) and final ($\psi_f$)
states and the dipole operator
${\bf A \cdot p}$ in the matrix elements of Eq. \ref{goldenrule}. 
These selection rules are especially relevant along the mirror and
glide planes, the directions $\overline\Sigma$ and $\overline\Lambda$
 in Fig. \ref{k_kaavio}. In the following analysis the dipole
 approximation is used, but the 
derivations are exact beoynd the approximation if the wave vector of
the radiation lies in the symmetry plane that is under discussion.

In the mirror plane, since there is a well defined parity about the
reflection in the plane, the
selection rules can be derived by considering the
parities of the components of the matrix element.\cite{Her76}
The transition is allowed
if the entire dipole matrix element $\langle \psi_f |{\bf A \cdot p}
|\psi_i\rangle$ is even. 
With respect to reflection the final state
with its momentum lying in the mirror plane is even. The operator
${\bf A \cdot p}$ is even(odd) if the polarization is parallel(perpendicular) to the
plane, which can be explained by considering parities of the initial
state $\psi_i$ and its derivative ${\bf A \cdot p}\psi_i$, where
$\boldsymbol p = -i\hbar\nabla$. The initial state at the Fermi level is similar to a
Bloch sum of copper $d_{x^2-y^2}$ orbitals\footnote{In the rotated
  orthorhombic coordinate system of Fig. \ref{Bilayer} these are $d_{xy}$ orbitals}. With respect to reflection
this orbital and the Bloch sum with its momentum lying in the plane is
odd.
Thus, for photointensity to
be nonzero the operator ${\bf A \cdot p}$ must be
odd and respectively polarization must be perpendicular to the mirror
plane. This result is presented in combined form in table \ref{saannot}.
The
even(identity) representation is denoted by $\Sigma$ and the odd representation
by $\Sigma'$.

Along the glide plane in the
orth. structure 
the full machinery of the group theory has to be used.
Photoemission selection rules due to glide plane have been considered previously
by Pescia \emph{et al.}\cite{Pes85}
A good account of a more general procedure has been given by Bassani
\emph{et al.}\cite{Bas75}.
The dipole matrix element $\langle \psi_f^{(\mu)} |({\bf A \cdot p})^{(\alpha)}
|\psi_i^{(\nu)}\rangle$ will vanish if the irreducible representation(irrep)
of
the final state does not appear in the product of irreps of
the initial state and the operator. 
The number of times it appears is given by
\cite{Bas75}
\begin{align}\label{saanto}
& c({\bf k},\mu;\alpha;{\bf k},\nu) = \nonumber \\ &\frac{1}{h_{{\bf k}}}
  \sum_{\{R_{{\bf k}}|{\bf f}\}} 
\chi^{(\mu)}(\{R_{{\bf k}}|{\bf f}\})^*
\chi^{(\alpha)}(R_{\boldsymbol k})
\chi^{(\nu)}(\{R_{{\bf k}}|{\bf f}\}),
\end{align}
where h$_{{\bf k}}$ is the order(number of elements $\{R_{\bf k}|{\bf f}\}$) of 
the relevant little group
for a particular $\boldsymbol k$-point in the first Brillouin zone.
$\chi^{(\nu)}(\{R_{{\bf k}}|{\bf f}\})$ is the character of
the little group irrep $\boldsymbol D^{(\nu)}(\{R_{{\bf k}}|{\bf
f}\})$ for the initial state, $\chi^{(\mu)}(\{R_{{\bf k}}|{\bf
  f}\})$ the character of the irrep of the final state 
and $\chi^{(\alpha)}(R)$ is the character of the
small point group representation $\boldsymbol D^{(\alpha)}(R_{\boldsymbol k})$ of the
dipole operator. 
The little group of a particular $k$ point contains operations
$\{R|{\bf f}\}$ of the full space group
that satisfy 
\begin{align}\label{lgroupk}
\boldsymbol R{\bf k} = {\bf k} + {\bf G},
\end{align}
i.e, leaving $\boldsymbol k$ unchanged. 
The small point group of ${\bf
  k}$ contains the rotational parts of the little group
operations. 

In the glide plane determined by direction $\overline\Lambda$ in
Fig. \ref{k_kaavio}, 
the little group(operations that 
keep $\boldsymbol k_\parallel = k_y{\bf
 \hat k_y}$ and $\boldsymbol k_z$
unchanged) of the $\boldsymbol k$ vector in the plane contains
two operations, identity operation $\{E|0\}$, and
glide operation $\{\sigma_{x}|{\bf b}/2\}$.
Both of the two
operations form a class of their own 
and the eigenstates in the plane can be classified by
two irreps.
Character table for the little group is in
table \ref{karakterit}.
\begin{table}[h]
  \caption{Character table of the little group in plane $\bar\Lambda$. $\delta$ is $e^{-i{\bf k \cdot b/2}}$.}
  \label{karakterit}
  \begin{center}
    \small \begin{tabular}{lcc}
      \hline\hline
      &    $E$    & $\{\sigma_{x}|{\bf b}/2\}$ \\
      \hline
      \hspace{0.0cm}$\Lambda$ \hspace{0.3cm} &\hspace{0.3cm} 1
      \hspace{0.3cm}        &  \hspace{0.3cm}$\delta$ \hspace{0.3cm}
      \\
      $\Lambda'$ \hspace{0.3cm}&  1        & -$\delta$ \\
      \hline\hline
      \end{tabular}
  \end{center}
\end{table}

To proceed
it has to be found out 
how to assign symmetry labels to
particular final (and initial) states. 
The matching of the wave function at the surface of the sample
rules out some of the possible final states. 
The final state in ARUPS is a
time-reversed LEED state\cite{Fei74}. The irrep of this state
may be obtained from the plane wave state $\psi_{\boldsymbol k_f}
= e^{i{({\bf k}_f
\cdot {\bf r})}}$ that arrives to the detector\cite{Her76}. 
The state is transformed under the glide operation
according to
\begin{align}
O&_{\{\sigma_{x}|{\bf b}/2\}}\psi_{{\bf k}_f}({\bf r}) 
 = \psi_{{\bf k}_f}(-x,y-b/2,z) \nonumber \\
&= e^{i(k_{fy}(y-b/2)+k_{fz}z)}
 = e^{-ik_{fy}b/2}\psi_{\boldsymbol k_f}({\bf r}) \nonumber \\
& = e^{-i(k_y +G_y)b/2}\psi_{\boldsymbol k_f}({\bf r})
 =\delta e^{-in\pi}\psi_{\boldsymbol k_f}({\bf r}),
\end{align}
where $k_y$ lies in the first Brillouin zone and $\boldsymbol G$ is a
reciprocal lattice vector.
The glide operation is represented by $ \delta e^{-in\pi}$,
where $n$ is integer. The irrep of the final state is a
function of the magnitude of the parallel component of its wavevector. It belongs to the irrep $\Lambda$ in the 1.,3.,5.,... repeated Brillouin zone,
and to the representation  $\Lambda'$ in the 2.,4.,6.,... zone.
This fundamental result for ARUPS and glide plane was first shown by
Pescia \emph{et al.}\cite{Pes85} and has been exploited for
Ni(100)-p(2x2)C\cite{Pri86} and TiO$_2$\cite{Har94}. 

Operator ${\bf A\cdot p}$
belongs to representation $\Lambda$ of the small point group if
polarization vector lies in the glide plane and to representation
$\Lambda'$ if it is perpendicular to the plane (parallel to
$x$-axis).
Photoemission selection rules follow from Eq. \ref{saanto}. They are
presented in table \ref{saannot}.
In summary the initial
state must belong to the same irrep as the final state if the
polarization is parallel to the symmetry plane and to the other
irrep if the polarization is perpendicular to the plane.

\begin{table}[h]
  \caption{Dipole selection rules in the planes of high symmetry.}
  \label{saannot}
  \begin{center}
    \small \begin{tabular}{clllllll}
\hline\hline
\multicolumn{7}{c}{$\boldsymbol k_f$ in the mirror plane $\overline\Sigma$} \\
\hline
Repeated zone of $\boldsymbol k_{f\parallel}$ & ... & 1. & 2. & 3. & 4. & 5. & ...\\
Allowed final State irrep   & ... & $\Sigma$  & $\Sigma$   &  $\Sigma$  & $\Sigma$
 & $\Sigma$  & ...\\
Allowed initial state if $\boldsymbol A\perp\overline\Sigma$ & ... & $\Sigma'$
&  $\Sigma'$ & $\Sigma'$  & $\Sigma'$ & $\Sigma'$ & ...\\
Allowed initial state if $\boldsymbol A\parallel\overline\Sigma$ & ... & $\Sigma$
&  $\Sigma$ & $\Sigma$  & $\Sigma$ & $\Sigma$ & ... \\
\hline\hline
\multicolumn{2}{l}{} \\
\hline\hline
\multicolumn{7}{c}{$\boldsymbol k_f$ in the glide plane $\overline\Lambda$} \\
      \hline
Repeated zone of $\boldsymbol k_{f\parallel}$ & ... & 1. & 2. & 3. & 4. & 5. & ...\\
Allowed final State irrep   & ... & $\Lambda$  &  $\Lambda'$ & $\Lambda$  &
$\Lambda'$ & $\Lambda$ & ...\\
Allowed initial state if $\boldsymbol A\perp\overline\Lambda$ 
 & ... & $\Lambda'$
&  $\Lambda$ & $\Lambda'$  & $\Lambda$ & $\Lambda'$ & ... \\
Allowed initial state if $\boldsymbol A\parallel \overline\Lambda$
& ... & $\Lambda$
&  $\Lambda'$ & $\Lambda$  & $\Lambda'$ & $\Lambda$ & ...\\
\hline\hline
      \end{tabular}
  \end{center}
\end{table}

To finish the symmetry labels of the initial states
forming FS have to be determined.
As was discussed the movements of the atoms from the tetr. sites are very
small in the copper-oxide layers and it can be assumed that the shape of the
initial state is very close to the tetr. case, which can also be shown in the band structure
calculations. Thus the initial state at FS belongs to the same irrep
as a Bloch state of copper d$_{x^2-y^2}$
orbitals. A Bloch sum of a Copper
$d_{x^2-y^2}$ orbitals at a
position
 ${\bf  d}_1 = 0.251{\bf a}+0.5{\bf b}-0.303{\bf c}$ is defined as
\begin{align}\label{blochsumma}
\phi_{\boldsymbol d_{x^2-y^2},Cu_1}(\boldsymbol r) = \sum_{\boldsymbol\tau_\nu }
e^{i{\bf k}\cdot \boldsymbol \tau_\nu}
\varphi_{d_{x^2-y^2}}({\bf r-d_1}-\boldsymbol \tau_\nu).
\end{align}
This can be decomposed into components
belonging to different irreducible
representations of the little group of $\boldsymbol k$ by\cite{Bas75}
\begin{align}
\phi_{\boldsymbol d_{x^2-y^2}}^{(\Lambda)}(\boldsymbol r) = 
& \frac{1}{2}\left(\phi_{\boldsymbol d_{x^2-y^2},Cu_1}(\boldsymbol r)-e^{-i{\bf
      k}\cdot {\bf b}/2}\phi_{\boldsymbol d_{x^2-y^2},Cu_2}(\boldsymbol r)\right), \nonumber\\ 
\phi_{\boldsymbol d_{x^2-y^2}}^{(\Lambda')}(\boldsymbol r) = 
&\frac{1}{2}\left(\phi_{\boldsymbol d_{x^2-y^2},Cu_1}(\boldsymbol r)+e^{-i{\bf
      k}\cdot {\bf b}/2}\phi_{\boldsymbol d_{x^2-y^2},Cu_2}(\boldsymbol r)\right),
\end{align}
where $\phi_{\boldsymbol d_{x^2-y^2},Cu_2}(\boldsymbol r)$ is a
Bloch sum of an identical $d_{x^2-y^2}$ orbital at ${\bf d}_2 = -0.251{\bf a}+0{\bf b}-0.303{\bf c}$.
\begin{figure}[]
  \subfigure[]{
    \resizebox{4.cm}{!}{\includegraphics{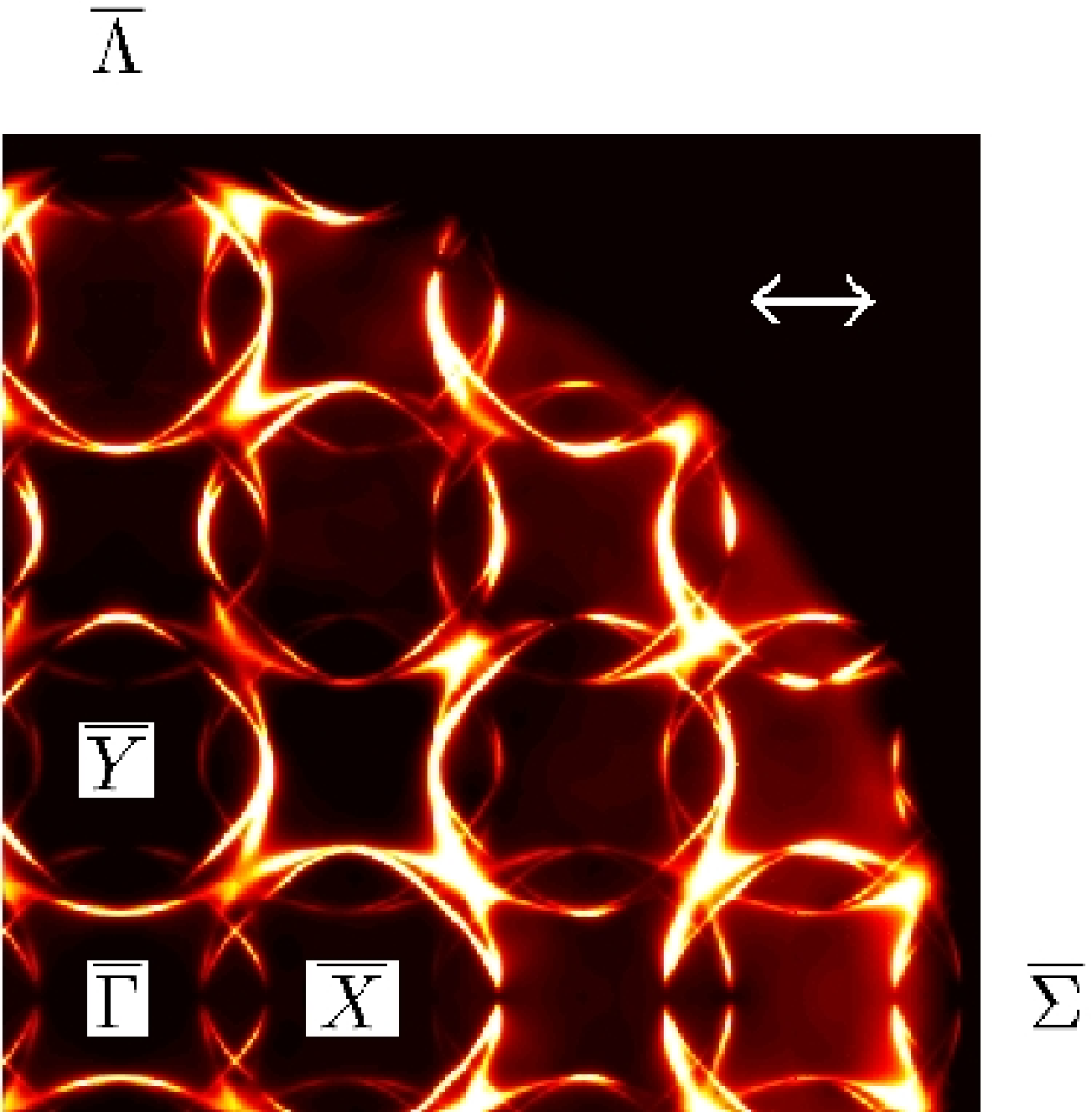}}
    \label{fermi00}
  }
  \subfigure[]{
    \resizebox{4.cm}{!}{\includegraphics{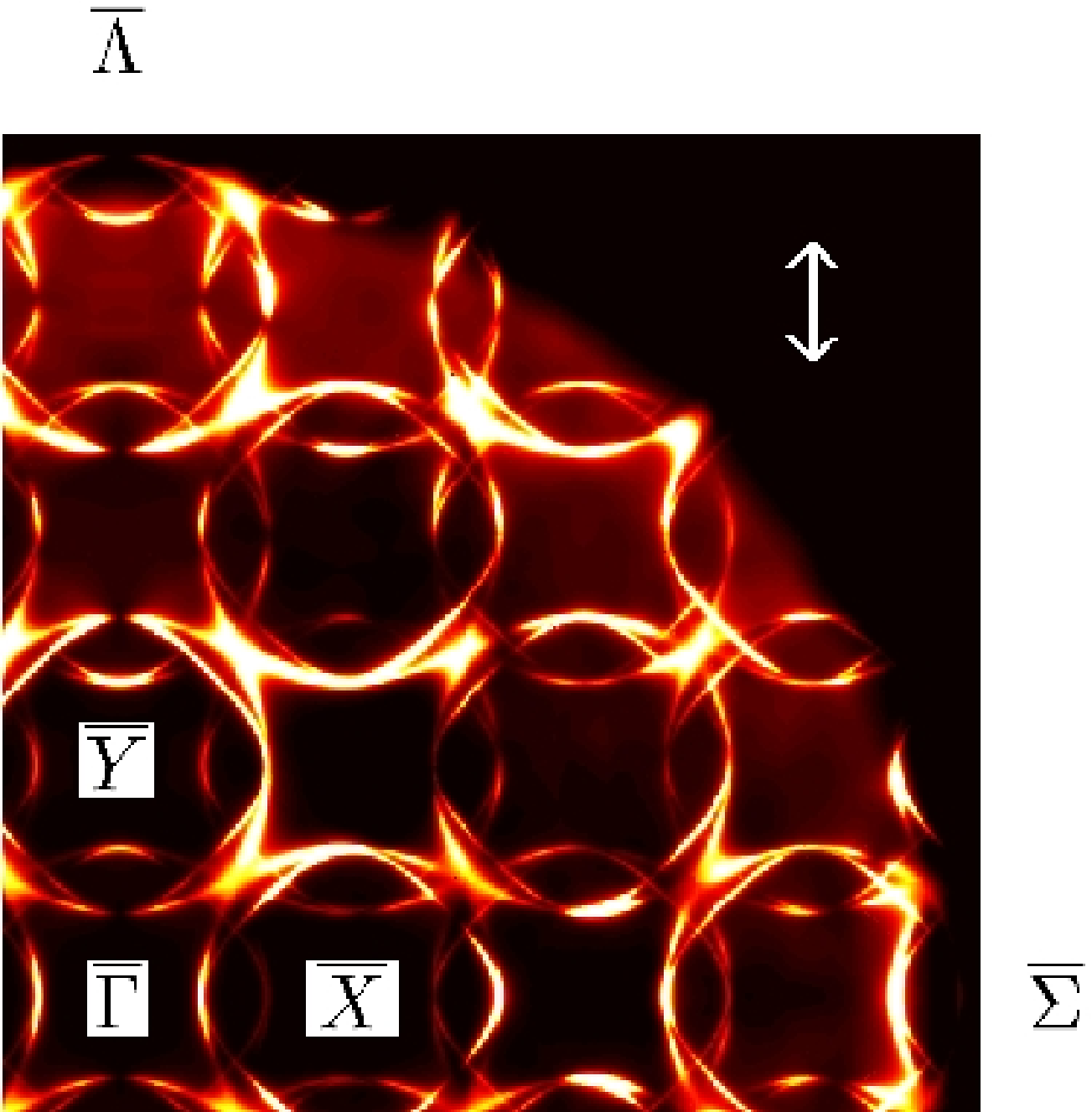}}
    \label{fermi90}
  }
\caption{ Bi2212 FS map. Normal incidence. (a) Polarization parallel
   to $x$-axis.  (b) Polarization parallel
   to $y$-axis.}
\label{isofermi00}
\end{figure}
These states can be pictured at $\Gamma$ as bonding-antibonding states which have the
opposite dispersion as a function of ${\bf k}_y$. 
$\phi_{d_{x^2-y^2}}^{(\Lambda')}$ is the bonding state with lower
energy and  $\phi_{d_{x^2-y^2}}^{(\Lambda)}$ the antibonding state.
In the repeated zone scheme the state $\phi_{\boldsymbol d_{x^2-y^2}}^{(\Lambda')}$  represents the main band i.e. the band
of the tetr. structure and the state $\phi_{\boldsymbol
  d_{x^2-y^2}}^{(\Lambda)}$ the shadow band.

We have performed first-principles simulations of ARUPS spectra in
Bi2212 using the one-step model of photoemission\cite{Pen76,Ban98}. 
Fig. \ref{isofermi00} shows the calculated
FS of the orth. Bi2212 in the normal incidence
setup. Double headed arrows indicate polarization of the incident
light. Photon energy was 54 eV. Firstly we would like to remark that there
seems to be no general correspondence between the intensitities of the main band
and the shadow band. The
effect of selection rules is seen in high symmetry directions
$\overline\Sigma$ and $\overline\Lambda$. The FS map is probed at the fermi
level and the initial state belongs to representation $\Lambda'$
in the glide plane and to the representation $\Sigma'$ in
the mirror plane.  In Fig. \ref{fermi00} polarization
is parallel to the mirror plane and neither the shadow or the main is
visible in $\overline\Sigma$-direction, whereas only the main band is visible in $\overline\Lambda$-direction. In Fig. \ref{fermi90} polarization
is parallel to the glide plane and both of the bands are
visible in $\overline\Sigma$-direction, whereas only the shadow band is
visible in $\overline\Lambda$-direction.
This phenomena has also been shown experimentally\cite{Man06,Izq05}. The strange behavior of
the shadow band follows from the variating irrep of the final
state in the consequent Brillouin zones.
\begin{figure}[]
      \resizebox{8.8cm}{!}{\includegraphics{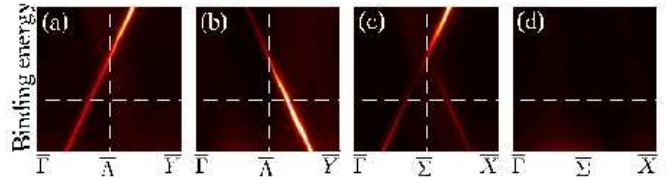}}
      \caption{Photoelectron spectra along high symmetry
	lines as a function of energy. (a) Along the glide plane,
	perpendicular polarization. (b) Along the glide plane, parallel
	polarization. (c) Along the mirror plane, parallel polarization.
	(d) Along the mirror plane, perpendicular polarization.}\label{edc}
\end{figure}

Fig. \ref{edc} shows photointensity as a function of binding energy
and $k_{\parallel}$. Calculations were performed within the normal incidence
setup, photon energy was 40 eV. The Fermi
function was ignored. Fig.
\ref{edc}(a) and \ref{edc}(b) show results along the glide plane with
polarization perpendicular to the plane(Fig.(a), main band visible)
and parallel to the plane(Fig.(b), shadow band visible). Fig.
\ref{edc}(c) and \ref{edc}(d) show the corresponding results along the
mirror plane.
Along the glide plane the main and the shadow band seem to give continuous
intencity as a function of $\boldsymbol k_\parallel$ even when crossing the
zone barrier. This can again be explained with table \ref{saannot}.
When an initial state band crosses the zone barrier it can be mapped
to the other band in the first Brillouin zone and it's 
irrep, as the irrep of the final state, will
change.

Furthermore, 
there is one surprising consequence of the selection rule due to glide
plane. As this rule states the final states for the shadow and the main band
must belong to a different irrep along
$\overline\Lambda$-line. Consequently, when picturing the band structure,
the final state for the shadow band cannot be an umklapp
of the final state for the main band, but on the contrary
it must lie in some other available band with fixed final state energy.
This means that the intensities of the bands varie e.g. as a function
of photon energy in an
uncorrelated way, which can also be seen as a function of binding
energy in Fig.\ref{edc}(a) and
\ref{edc}(c). Actually because of the fact that the width (due to
$\Sigma_f''$ and $\boldsymbol k_\perp$ dispersion\cite{Sah05})
of the final state is relatively large, both the main and
the shadow band can be observed with the same single photon energy.

Available experimental data are consistent with our predicted
intensities due to structural distortions, but we also point out that
the intensity variation along the glide plane strongly masks any weak sign
of antiferromagnetism(AFM) induced 
spectral features.
We have modelled the molecular field of the AFM interaction with planar ordering, which is known to exist in La$_2$CuO$_{4-y}$\cite{Vak87}, by an \emph{ad hoc} parameter $u$ on copper sites. Our computations
reveal that, to produce shadow bands in ARUPS by AFM,
relatively large $u$ has to be used.

\begin{figure}[!t]
      \resizebox{8.8cm}{!}{\includegraphics{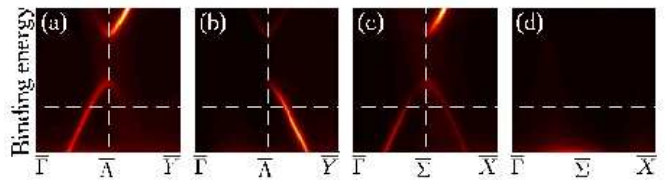}}
\caption{Photoelectron spectra along the high symmetry
    lines as a function of energy, including AFM-effect, $u=$ 0.9 eV. (a) Along the
glide plane,
      perpendicular polarization. (b) Along the glide plane, parallel
      polarization, (c) Along the mirror plane, perpendicular polarization,
      (d) Along the mirror plane, parallel polarization.}\label{afmdc}
\end{figure}

But there is one interesting possiblitity. If both the
structural and the AFM interactions were present, as must be with low
doping, the selection rules
would change again. The correct magnetic group should be determined,
but because the specification of the magnetic moments in Bi2212 is not
available and because of limitations of our program code, we can only discuss
the AFM selection rules qualitatively.  
While maintaining the reflection symmetry the \emph{ad hoc} planar 
ordering breaks the glide symmetry, and, consequently, the
derived selection rules along the glide
plane would be no longer valid. In practice, this would mean that, in those high symmetry directions where the
structural distortions forbid intensity,
there could exist signatures of the FS due to the AFM interaction. 
In Fig. \ref{afmdc} we have calculated
the effect of the AFM on ARUPS spectra. The
magnitude of $u$ was 0.9 eV, which induces an energy gap of the same
magnitude.  In contrast to Fig. \ref{edc}, along the glide plane
(Fig. \ref{afmdc}a and Fig. \ref{afmdc}b), there are no longer strict
selection rules for the polarization, and the main and the shadow band
are, though weakly, visible
with both polarizations. Along the mirror plane [Figs. \ref{afmdc}c and
\ref{afmdc}d] the selection
rules remain strict.
We thus urge experimetalists to
focus some measurements with high flux and precise polarizations to look possible signatures of
AFM features. A development of the AFM gap with lowered doping and
maybe other signatures of the AFM may
be seen in this way.

We have shown that structural modifications in Bi2212 from a body
centered tetragonal lattice to a base-centered 
orthorhombic lattice change selection rules in ARUPS along the high symmetry lines.
Because of the glide plane in orthorhombic structure, it turns out that the irreducible representation of the final state changes
alternatively in the repeated zone scheme of $\boldsymbol k_\parallel$ space. With fixed polarization this yields opposite
intensity behaviour for the main and the shadow FS in the adjacent Brillouin
zones. The calculated FS is consistent with available
experimental data, which is strong evidence against the
antiferromagnetic scenario, but a possibility of detecting AFM features in
low doping is proposed.

\begin{acknowledgments}

   This work is funded from the Academy of Finland and benefited from
   the Institute of Advanced Computing (IAC), Tampere.

\end{acknowledgments}

\bibliographystyle{apsrev}
\bibliography{references}

\end{document}